\documentclass[sigconf]{acmart}

\usepackage{algorithmic}
\usepackage{textcomp}
\usepackage{xcolor}
\usepackage{microtype}
\usepackage{fontawesome5}
\usepackage[inline]{enumitem}
\usepackage{multirow}
\usepackage{pifont}

\setcopyright{none}
\copyrightyear{2026}
\acmYear{2026}
\acmDOI{XXXXXXX.XXXXXXX}
\acmConference[TechDebt 2026]{International Conference on Technical Debt}{April 12--15, 2026}{Rio de Janeiro, Brazil}

\begin{document}

\title{Technical Lag as Latent Technical Debt: A Rapid Review}

\author{Shane K. Panter}
\email{shanepanter@boisestate.edu}
\orcid{0009{-}0009{-}1852{-}3655}
\affiliation{%
  \institution{Boise State University}
  \city{Boise}
  \state{Idaho}
  \country{USA}
}

\author{Nasir U. Eisty}
\email{neisty@utk.edu}
\orcid{0000{-}0001{-}5228{-}4664}
\affiliation{%
  \institution{University of Tennessee}
  \city{Knoxville}
  \state{Tennessee}
  \country{USA}
}

\renewcommand{\shortauthors}{Panter and Eisty}

\begin{abstract}
\textit{Context:} Technical lag accumulates when software systems fail to keep pace with technological advancements, leading to software quality deterioration. \textit{Objective:} This paper aims to consolidate existing research on technical lag, clarify definitions, explore its detection and quantification methods, examine underlying causes and consequences, review current management practices, and lay out a vision as an indicator of passively accumulated technical debt. \textit{Method:} We conducted a Rapid Review with snowballing to select the appropriate peer-reviewed studies. We leveraged the ACM Digital Library, IEEE Xplore, Scopus, and Springer as our primary source databases. \textit{Results:} Technical lag accumulates passively, often unnoticed due to inadequate detection metrics and tools. It negatively impacts software quality through outdated dependencies, obsolete APIs, unsupported platforms, and aging infrastructure. Strategies to manage technical lag primarily involve automated dependency updates, continuous integration processes, and regular auditing. \textit{Conclusions:} Enhancing and extending the current standardized metrics, detection methods, and empirical studies to use technical lag as an indication of accumulated latent debt can greatly improve the process of maintaining large codebases that are heavily dependent on external packages. We have identified the research gaps and outlined a future vision for researchers and practitioners to explore.
\end{abstract}



\received{10 October 2025}
\received[accepted]{5 January 2026}

\maketitle

\section{Introduction}\label{sec:introduction}

Gonzalez-Barahona et al.~\cite{gonzalez-barahonaTechnicalLagSoftware2017} coined the term ``technical lag'' (TL) to recognize the differences between technical debt (TD), which is the result of deliberate trade-offs in software design and implementation~\cite{cunninghamWyCashPortfolioManagement1992}, and the impact of outdated dependencies that are used in the deployment process. TL accumulates passively rather than through intentional design decisions and measures how an ecosystem ``degrades'' with just the passing of time~\cite{zeroualiEmpiricalAnalysisTechnical2018, zeroualiFormalFrameworkMeasuring2019}. Both TL and TD can be challenging to detect or quantify without the use of appropriate tools and metrics~\cite{melinExploringAdvancesUsing2025, panterPVACPackageVersion2025}. Additionally, as demonstrated in the recent \texttt{XZ Utils}\footnote{\href{https://nvd.nist.gov/vuln/detail/CVE-2024-3094}{https://nvd.nist.gov/vuln/detail/CVE-2024-3094}} supply chain incident and research on \texttt{npm} packages~\cite{zeroualiEmpiricalAnalysisTechnical2018}, driving TL to zero by running bleeding-edge software also carries risk by introducing components that may not have been thoroughly vetted.

The prevalence of TL is not new and is alarmingly widespread. By establishing a clearer connection between these two constructs, we aim to position TL not merely as a maintenance nuisance, but as a hidden dimension of TD and a potential indicator of latent technical debt (LTD). Studies demonstrate that one out of four dependencies and two out of five releases in the \texttt{npm} ecosystem suffer from TL~\cite{decanEvolutionTechnicalLag2018}. Operating system package managers may sometimes be outdated for months or even years~\cite{legay_quantitative_2021, panterPVACPackageVersion2025}. This difference between the upstream source and what is being used significantly elevates the risk of vulnerability exposure relative to regularly maintained components~\cite{cox_measuring_2015, zeroualiRelationOutdatedDocker2019}.

To address these challenges, researchers and practitioners have increasingly emphasized the need for comprehensive approaches to detect, quantify, and mitigate TL~\cite{gonzalez-barahonaCharacterizingOutdatenessTechnical2020, zeroualiFormalFrameworkMeasuring2019, panterPVACPackageVersion2025}. However, despite significant contributions in recent years, there is still much work to be done regarding the application of metrics, the development of robust empirical methodologies, and the exploration of broader ecosystems. Additionally, prior studies quantify the time to update, but not the safety of updating; there is no metric for ``safe to adopt'' time under supply-chain compromise.

By consolidating existing definitions, processes, empirical findings, and management strategies, we provide a comprehensive synthesis that clarifies the scope of TL, examines its root causes and consequences, evaluates the effectiveness of current detection and mitigation practices, and identifies gaps and opportunities for future research. Through this Rapid Review synthesis, we contribute essential foundations for future research and practical approaches to improve software sustainability, reliability, and innovation.

The contributions of this paper are:

\begin{enumerate}
    \item Consolidating and categorizing metrics and methods for detecting TL.\@
    \item Identifying five key research gaps across ecosystems.
    \item Establishing the conceptual link between TL and LTD.\@
\end{enumerate}

\section{Background}\label{sec:background}

TL and TD share common consequences (Table~\ref{tab:debt-vs-lag}), including increased maintenance costs, reduced developer productivity, and elevated security risks. For example, studies consistently demonstrate that outdated dependencies are significantly more likely to contain vulnerabilities and defects~\cite{cox_measuring_2015, huEmpiricalAnalysisVulnerabilities2024, zeroualiImpactOutdatedVulnerable2019}. These risks parallel those described in the TD literature, where unaddressed shortcuts manifest as higher defect rates and system fragility over time~\cite{brownManagingTechnicalDebt2010, edbertExploringTechnicalDebt2023, zazworkaInvestigatingImpactDesign2011}. Thus, TL can be seen as a ``debt-like liability'' incurred not by intentional shortcuts, but by inaction and inertia in dependency management. Fig.~\ref{fig:tech-lag-debt} illustrates the relationship between TL and TD.\@

\begin{figure*}[hbtp]
    \centering
    \includegraphics[width=1\linewidth]{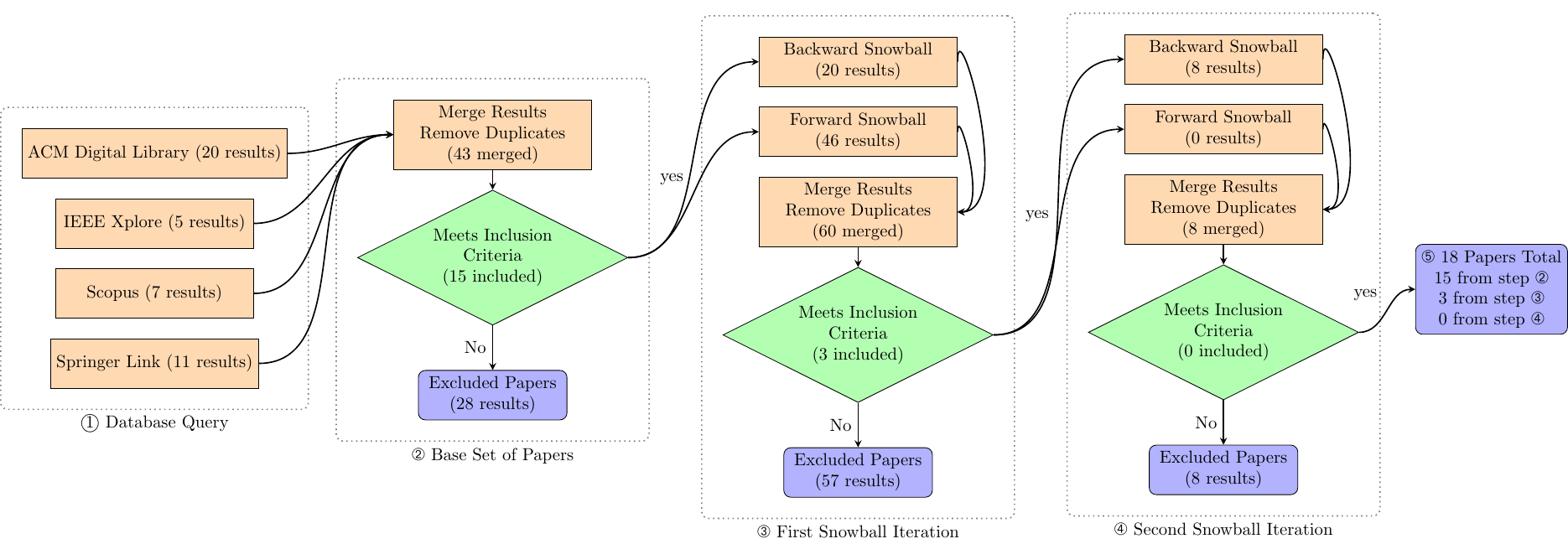}
    \caption{Results from the initial database query \ding{192} were filtered by our inclusion and exclusion criteria \ding{193}. Two rounds of forward and backward snowballing were completed \ding{194}, \ding{195} to yield a final set of studies.}\label{fig:searchStrategy}
    \Description{A flowchart showing the search strategy for the Rapid Review. The first box shows the initial database query yielding 43 unique papers. An arrow points to the second box, which shows filtering by inclusion and exclusion criteria, resulting in 15 papers. Two arrows point from this box to two additional boxes, representing forward and backward snowballing, which add 3 more papers, resulting in a final set of 18 studies.}
\end{figure*}

TL introduces a form of compound interest in the repayment of TD.\@ When dependencies are left outdated, their eventual update not only requires larger effort due to version incompatibilities but also forces developers to refactor dependent modules, adjust tests, and retrain teams on new APIs. This compounding effect aligns closely with the ``interest'' metaphor in TD, where the longer the debt is left unaddressed, the more costly it becomes to repay.

\section{Goal and Research Questions}\label{sec:researchQuestions}

The goal of our study is to establish the current state of TL research with a Rapid Review and establish a future vision for identifying LTD.\@ We aim to identify all the metrics that have been created, the ecosystems studied, and the research gaps that exist.

Based on this goal, our research questions are as follows:

\begin{description}
    \item[RQ1:] What are the primary causes of technical lag in software projects?
    \item[RQ2:] What metrics and methods have been proposed to detect or quantify technical lag?
    \item[RQ3:] What are the research gaps in understanding or addressing technical lag?

\end{description}

\section{Search Strategy}\label{sec:researchMethodology}

We conducted a Rapid Review following Cartaxo et al.~\cite{cartaxoRoleRapidReviews2018} and Wohlin's snowballing guidelines~\cite{wohlinGuidelinesSnowballingSystematic2014} to synthesize existing research on TL.\@ Searches were conducted in the ACM Digital Library, IEEE Xplore, Scopus, and Springer. We included peer-reviewed studies published between 2015 and 2025 that explicitly addressed TL in software ecosystems and answered at least one of our research questions. Initial queries identified 43 unique papers, of which 15 met the inclusion criteria. Snowballing added 3 more, resulting in a final dataset of 18 studies (Fig.~\ref{fig:searchStrategy}). \textbf{The complete dataset has been made available for replication~\cite{panterTechnicalLagLatent2026}.}

The final search strings are as follows:
\begin{description}
    \item[ACM] ``query'': {``Technical Lag''} ``filter'': \{Article Type: Research Article\}, \{ACM Content: DL\}
    \item[IEEE] ``All Metadata'': ``Technical Lag''
    \item[Scopus] TITLE-ABS-KEY (``Technical Lag'') AND (LIMIT-TO (SUBJAREA, ``COMP'')) AND (LIMIT-TO (DOCTYPE, ``ar'')) OR (LIMIT-TO DOCTYPE, ``cp'')
    \item[Springer] Keywords: ``Technical Lag'' Content Type: (Research Article, Conference Paper) Disciplines: (Computer Science)
\end{description}

\subsection{Selection Procedure}\label{sec:inclusionExclusionCriteria}

The first author performed each step detailed in Fig.~\ref{fig:searchStrategy} while the second author confirmed the results. Any disagreements were resolved through discussion. In steps \ding{192}, \ding{193}, and \ding{194}, papers were screened in two rounds. The first round screened each potential paper based on its title and abstract. The second round consisted of a full-text review to select the most relevant studies.

\subsection{Extraction Procedure}

Data relevant to our research questions were extracted from each paper and grouped by research question by the first author for analysis and synthesis. The second author reviewed the results to ensure all relevant data were correctly identified.

\section{Results}\label{sec:dataSynthesis}

In this section, we present the results of our Rapid Review and answer the research questions posed in Section~\ref{sec:researchQuestions}.

\subsection{\textbf{RQ1: Primary Causes}}\label{sec:rq1-results}

TL in software projects arises from a mix of rigid dependency policies, slow update practices, ecosystem-specific requirements, and tooling limitations~\cite{zeroualiEmpiricalAnalysisTechnical2018, stringerTechnicalLagDependencies2020, legay_quantitative_2021, panterPVACPackageVersion2025}. Studies in the \textit{npm} ecosystem report that strict version constraints and fixed declarations are a significant cause of TL, where the typical TL is 7 to 9 months, with 25\% having a TL of more than 9 months and only 25\% with a TL less than 52 days~\cite{decanEvolutionTechnicalLag2018}.

These findings underscore that TL is not a single factor but rather the combined effect of ecosystem factors, developer decision patterns, and infrastructure and tooling challenges~\cite{legay_quantitative_2021, zeroualiFormalFrameworkMeasuring2019, panterPVACPackageVersion2025, decanOutdatednessWorkflowsGitHub2023, opdebeeckDockerHubImage2023, heAutomatingDependencyUpdates2023}.

\subsubsection{\textbf{Ecosystem Factors}}
Software ecosystems exhibit distinct characteristics significantly influencing how TL manifests and evolves. Different package managers and distribution platforms have developed unique approaches to managing dependencies and updates, leading to varying outcomes in terms of TL~\cite{panterPVACPackageVersion2025}. Packages can be provided by the language ecosystem (e.g., \texttt{npm}, \texttt{Maven}), operating system distributions (e.g., \texttt{apt}, \texttt{yum})~\cite{panterPVACPackageVersion2025,legay_quantitative_2021}, or container registries (e.g., \texttt{Docker Hub})~\cite{zeroualiRelationOutdatedDocker2019}, making it difficult to get a complete picture.

In the Linux distribution landscape, contrasting philosophies drive update patterns. Arch Linux prioritizes freshness in its package management, resulting in lower TL across its ecosystem. In stark contrast, CentOS takes a more conservative approach, emphasizing stability over currency, which naturally leads to a higher TL but potentially more stable systems~\cite{legay_quantitative_2021}. This approach makes direct comparisons difficult~\cite{panterPVACPackageVersion2025} when applications are deployed in multiple environments.

The \texttt{Maven}, \texttt{Docker}, and \texttt{npm}  ecosystems present their unique challenges. These platforms struggle particularly with transitive dependencies and the ongoing maintenance of images and packages~\cite{zeroualiFormalFrameworkMeasuring2019, zeroualiImpactOutdatedVulnerable2019, zeroualiAnalyzingTechnicalLag2018}. In Docker's case, there's a notable difference between official and community images, with nearly 70\% of popular child images inheriting an outdated parent with a median of 5.63 months, and community images having higher lag than officially maintained images~\cite{opdebeeckDockerHubImage2023}.

A common thread across ecosystems is the impact of dependency tree complexity~\cite{zeroualiFormalFrameworkMeasuring2019,panterPVACPackageVersion2025}. Deep dependency trees that span multiple package managers at both the language and operating system levels may not be accurate and can create additional package maintenance and update complications. This complexity is particularly evident in \texttt{npm}, where the depth and interconnectedness of dependencies can create significant bottlenecks in the update process~\cite{heAutomatingDependencyUpdates2023}.

\subsubsection{\textbf{Developer Decision Patterns}}
In the complex software development landscape, developers face ongoing decisions about managing their project dependencies. This leads to distinct patterns in how they approach updates and manage any potential breaking changes~\cite{zeroualiFormalFrameworkMeasuring2019, salzaThirdpartyLibrariesMobile2020}. These patterns reveal a delicate balance between maintaining stability and staying current with the latest versions.

Developers often find themselves at a crossroads between restrictive and permissive approaches when choosing update strategies. The consequences of these choices are significant; studies show that Java projects in the \texttt{Maven} ecosystem experience significant update delays~\cite{stringerTechnicalLagDependencies2020,cox_measuring_2015}. Fear plays a crucial role in these decisions. Developers frequently avoid updates due to concerns about breaking changes, prioritizing stability over currency. This cautious approach is further reinforced by the high costs associated with testing updates and developers' limited influence over their dependencies~\cite{decanEvolutionTechnicalLag2018, cox_measuring_2015, stringerTechnicalLagDependencies2020}.

\subsubsection{\textbf{Infrastructure and Tooling Impact}}

The infrastructure and tooling landscape is pivotal in managing and combating TL, yet current solutions present opportunities and significant challenges~\cite{heAutomatingDependencyUpdates2023, decanOutdatednessWorkflowsGitHub2023}. The effectiveness of these tools can make the difference between maintaining current dependencies and falling behind in critical updates~\cite{zeroualiFormalFrameworkMeasuring2019, huEmpiricalAnalysisVulnerabilities2024}.

At the foundation of modern dependency management are dependency graphs, which provide clear visibility into project dependencies. However, these crucial tools often fall short of their intended purpose~\cite{panterPVACPackageVersion2025}. Recent studies have revealed that dependency graphs can contain significant noise and increase workloads, potentially leading to undetected vulnerabilities~\cite{heAutomatingDependencyUpdates2023}. This unreliability in fundamental tooling creates a challenging foundation for dependency management.

\subsection{\textbf{RQ2: Metrics and Methods}}\label{sec:rq2-results}

Table~\ref{tab:metrics-crosswalk} maps each study to one of six categories of TL, the metrics used or defined, and the ecosystems studied. The categories are not mutually exclusive, as some studies proposed multiple metrics or modified existing ones to address ecosystem-specific challenges. The most common categories were Time Lag (10 studies) and Version Lag (6 studies), while Package Lag, Vulnerability and Bug Lag, Opportunity Lag, and Conceptual frameworks were less frequently addressed.

The TL categories are defined as follows:

\begin{enumerate}
    \item \textbf{Time Lag} measures the temporal gap between the release of a newer version and its adoption.
    \item \textbf{Version Lag} counts the number of releases newer than the latest installable and is optionally weighted by SemVer class (major.minor.patch).
    \item \textbf{Package Lag} focuses on the number or proportion of outdated packages within a system
    \item \textbf{Vulnerability and Bug Lag} assesses lag based on known vulnerabilities and bugs in dependencies, highlighting security risks associated with outdated components.
    \item \textbf{Opportunity Lag} measures the time window of active availability when updates were available but not adopted, emphasizing missed opportunities for improvement.
    \item \textbf{Conceptual} has defined theoretical aspects of technical lag that can be adopted to any ecosystem.
\end{enumerate}

\begin{table*}[htbp]
\footnotesize
\centering
\caption{Crosswalk of technical lag metric categories, concrete metrics, and ecosystems used in the literature.}
\begin{tabular}{p{1.8cm} p{8cm} p{3.3cm} p{3.3cm}}
\toprule
\textbf{Lag Category} & \textbf{Metric} & \textbf{Study} & \textbf{Ecosystem}\\
\midrule
 \textbf{Time}
 & Time$\Delta$ expressed in months between the release of an Action and the latest available release.  & Decan et al.~\cite{decanOutdatednessWorkflowsGitHub2023} & GitHub Actions (SEART GitHub Search Engine)\\
& $time\text{-}lag(\textit{i})$ per release time difference in packages  & Zerouali et al.~\cite{zeroualiMultidimensionalAnalysisTechnical2021}& Debian-based Docker images\\
 & Time$\Delta$ Delay in adopting newer library releases in apps (project history) & Salza et al.~\cite{salzaThirdpartyLibrariesMobile2020} & Android apps (ANDROIDTIMEMACHINE)\\
 & Time$\Delta$ calculated as the number of days (major, minor, and micro lag) & Stringer et al.~\cite{stringerTechnicalLagDependencies2020} & 14 package managers (libraries.io)\\
 & Time$\Delta$ date of first appearance of package version& Legay et al.~\cite{legay_quantitative_2021} & Arch, Debian, CentOS, Fedora, Ubuntu\\
 & $tLag(d)$ for a dependency package network & Zerouali et al.~\cite{zeroualiEmpiricalAnalysisTechnical2018} & npm (libraries.io)\\
 & Time$\Delta$ with modifications & He et al.~\cite{heAutomatingDependencyUpdates2023}  & GitHub repos (GHTorrent)\\
 & $\Delta_t(d,t)$ Lag of dependency $d$ at time $t$ & Decan et al.~\cite{decanEvolutionTechnicalLag2018} & npm (libraries.io)\\
 & Time$\Delta$ version release dates & Zerouali et al.~\cite{zeroualiImpactOutdatedVulnerable2019} & Docker Hub\\
 & Time$\Delta$ push date of latest parent vs.\@ inherited date & Opdebeeck et al.~\cite{opdebeeckDockerHubImage2023}  & Docker Hub\\

\midrule
\textbf{Version}

& $vers\text{-}lag(\textit{i})$ number of missed versions  & Zerouali et al.~\cite{zeroualiMultidimensionalAnalysisTechnical2021} & Debian-based Docker images\\
& Version$\Delta$ (major, minor and micro lag) & Stringer et al.~\cite{stringerTechnicalLagDependencies2020} & 14 package managers (libraries.io)\\
& $vLag(d)$ for a dependency package network & Zerouali et al.~\cite{zeroualiEmpiricalAnalysisTechnical2018} & npm (libraries.io)\\
& Version$\Delta$ package number of versions behind & Zerouali~\cite{zeroualiAnalyzingTechnicalLag2018} & Alpine-based Docker images\\
& Version$\Delta$ version behind latest & Zerouali et al.~\cite{zeroualiImpactOutdatedVulnerable2019} & Docker Hub\\
& Version Number Delta (VND) aggregated SemVer deltas & Panter et al.~\cite{panterPVACPackageVersion2025} & Debian/Ubuntu apt ecosystem (Ultimate Debian Database)\\
\midrule
\textbf{Package}

& $pkg\text{-}lag(\textit{i})$ number of outdated packages in the image &  Zerouali et al.~\cite{zeroualiMultidimensionalAnalysisTechnical2021} & Debian-based Docker images\\
& package time lag (in days) & Zerouali~\cite{zeroualiAnalyzingTechnicalLag2018} & Alpine-based Docker images\\
& Activity Categorizer (AC) Activity lag of Debian packages  & Panter et al.~\cite{panterPVACPackageVersion2025} & Debian/Ubuntu apt ecosystem (Ultimate Debian Database)\\

\midrule
\textbf{Vulnerability \& Bug}

& $vuln\text{-}lag(\textit{i})$ Count/severity of known CVEs present in the used versions vs what would be present if updated &  Zerouali et al.~\cite{zeroualiMultidimensionalAnalysisTechnical2021} & Debian-based Docker images\\
& Vulnerabilities and bugs in images (correlational analysis) & Zerouali et al.~\cite{zeroualiRelationOutdatedDocker2019} & Debian-based Docker images\\
& $bug\text{-}lag(\textit{i})$ Open/known bug counts persisting in the used versions vs updated baseline &  Zerouali et al.~\cite{zeroualiMultidimensionalAnalysisTechnical2021} & Debian-based Docker images\\
& Patch lagging time $T_{fix} (commit), T_{ver} (version), T_{index}(Index)$ & Hu et al.~\cite{huEmpiricalAnalysisVulnerabilities2024} & Golang\\

\midrule
\textbf{Opportunity}
& Total time an update has been available (opportunity window) but not adopted & Decan et al.~\cite{decanOutdatednessWorkflowsGitHub2023} & GitHub Actions (SEART GitHub Search Engine)\\
\midrule
\textbf{Conceptual}

& Theoretical model of Technical Lag & Gonzalez-Barahona et al.~\cite{gonzalez-barahonaTechnicalLagSoftware2017} & Debian Packages\\
& Time$\Delta$ and Version$\Delta$ formal definitions & Zerouali et al.~\cite{zeroualiFormalFrameworkMeasuring2019} & npm (libraries.io)\\
& Characterizing outdateness and technical lag & Gonzalez-Barahona et al.~\cite{gonzalez-barahonaCharacterizingOutdatenessTechnical2020} & Examples for language and OS package managers\\
& Introduced $Ver_{Sequence}$, $Ver_{Release}$, $Ver_{Delta}$ & Cox et al.~\cite{cox_measuring_2015} & Java (Maven)\\
\bottomrule
\end{tabular}\label{tab:metrics-crosswalk}
\end{table*}

\subsection{\textbf{RQ3: Research Gaps}}\label{sec:rq3-results}

TL appears to be widespread in package management ecosystems~\cite{zeroualiAnalyzingTechnicalLag2018, decanEvolutionTechnicalLag2018, panterPVACPackageVersion2025, legay_quantitative_2021}. Studies report that dependency updates are often delayed because existing metrics typically focused on time and version lag do not capture key factors such as update effort, security risks, ecosystem-specific practices~\cite{cox_measuring_2015, zeroualiEmpiricalAnalysisTechnical2018, legay_quantitative_2021, zeroualiMultidimensionalAnalysisTechnical2021} or lost opportunities~\cite{decanOutdatednessWorkflowsGitHub2023}. In \texttt{npm}, \texttt{Maven}, and \texttt{Docker} ecosystems, research indicates that TL grows over time due to transitive dependencies, versioning policies, and challenges in automated tooling~\cite{decanEvolutionTechnicalLag2018, zeroualiEmpiricalAnalysisTechnical2018, zeroualiAnalyzingTechnicalLag2018, stringerTechnicalLagDependencies2020, panterPVACPackageVersion2025}. Inaccuracies in data sources and dependency graphs further complicate measurement~\cite{heAutomatingDependencyUpdates2023,opdebeeckDockerHubImage2023}.

\subsubsection{\textbf{Research Gap 1 Methods and Metrics}}
Conceptual Frameworks have laid the groundwork for measuring TL.\@ However, each ecosystem presents unique challenges in applying these frameworks, which need to be refined to account for the differences in available metadata between ecosystems. The complexity of managing these dependencies is compounded by versioning policies and the need to balance update frequency with associated risks and effort~\cite{stringerTechnicalLagDependencies2020, heAutomatingDependencyUpdates2023}. Developing new methods and metrics that address specific ecosystem challenges is crucial for a comprehensive understanding of the system~\cite{panterPVACPackageVersion2025}.

\subsubsection{\textbf{Research Gap 2 Data Quality and Coverage}}
Most work focuses on a few ecosystems (Table~\ref{tab:metrics-crosswalk}), leaving many systems untouched. Real-world ecosystems often lack the necessary data to evaluate TL~\cite{opdebeeckDockerHubImage2023}, which is filtered out~\cite{panterPVACPackageVersion2025}, resulting in incomplete datasets and underscoring the need for improved data collection and validation in real-world systems. Additionally, studies have found significant variations in TL patterns based on package manager policies, distribution types, and community practices~\cite{zeroualiFormalFrameworkMeasuring2019, zeroualiAnalyzingTechnicalLag2018, legay_quantitative_2021,panterPVACPackageVersion2025}. Ecosystems such as \texttt{Chocolatey} for Windows, \texttt{Homebrew} for macOS,  and \texttt{vcpkg} for C/C++  remain unexplored. The lack of data on these ecosystems limits the understanding of TL in these environments.

\subsubsection{\textbf{Research Gap 3 Automation and Tooling}}
Although tools like Dependabot help mitigate lag, compatibility assessments and developer trust remain problematic~\cite{heAutomatingDependencyUpdates2023}. Tooling that works out of the box without extensive customization would improve the adoption of TL methods. Continuous Ecosystem-level monitoring to spot problematic packages early can help teams actively manage their TL before it becomes problematic~\cite{panterPVACPackageVersion2025}. Additionally, the automated tooling itself can suffer from TL due to the need for manual updates and maintenance. This can lead to outdated tools used to manage dependencies, creating a vicious cycle of TL~\cite{heAutomatingDependencyUpdates2023, decanOutdatednessWorkflowsGitHub2023}.

\subsubsection{\textbf{Research Gap 4 Security and Risk Integration}}
A gap remains in integrating security considerations into lag metrics and management tools. While some studies have explored the relationship between TL and security vulnerabilities, there is a need for more comprehensive frameworks that incorporate security risk assessments into lag management practices~\cite{cox_measuring_2015, huEmpiricalAnalysisVulnerabilities2024, zeroualiImpactOutdatedVulnerable2019, zeroualiRelationOutdatedDocker2019,salzaThirdpartyLibrariesMobile2020}. Additionally, TL metrics need to include data to distinguish between deliberate protective delay and debt accruing neglect to avoid potential supply-chain attacks.

\subsubsection{\textbf{Research Gap 5 Developer Guidelines and Predictive Modeling}}
Limited guidance exists for dependency management practices, and predictive models require greater sophistication. While some studies have proposed models for predicting TL, these models often lack the necessary complexity to account for the factors influencing lag accumulation~\cite{panterPVACPackageVersion2025}. More sophisticated models that consider the interplay of various factors, such as developer behavior, ecosystem policies, and tooling limitations, are needed to provide actionable insights for practitioners~\cite{zeroualiEmpiricalAnalysisTechnical2018, zeroualiAnalyzingTechnicalLag2018, decanEvolutionTechnicalLag2018}.

\section{Technical Lag as Latent Technical Debt}

\begin{figure}[ht]
    \centering
    \includegraphics[width=1\linewidth]{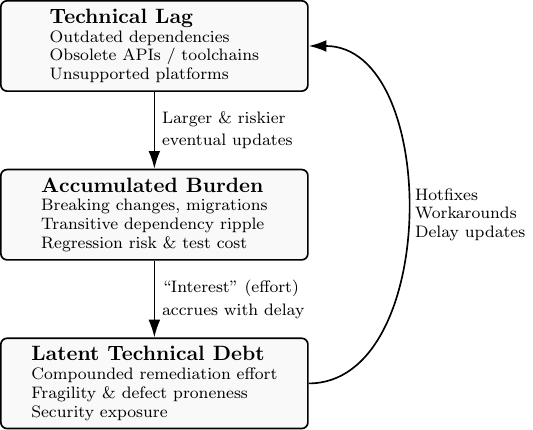}
    \caption{A Technical Lag feedback loop for Technical Debt.}\label{fig:tech-lag-debt}
    \Description{A diagram illustrating the feedback loop between Technical Lag and Technical Debt. The diagram shows that Technical Lag leads to increased maintenance effort, which in turn increases Technical Debt. This increased debt can lead to further Technical Lag, creating a continuous cycle.}
\end{figure}

TL manifests as a hidden liability and, as shown in Fig.~\ref{fig:tech-lag-debt}, can create a feedback loop where accumulated burden leads to LTD.\@ Inaction, hotfixes, and workarounds all contribute to deferred costs that resemble traditional debt, even though their origins differ.

Just as unresolved TD accrues ``interest'' in the form of additional maintenance effort, TL compounds the cost of eventual updates. Delayed upgrades often require larger migrations, adjustments to dependent modules, and extensive retesting. This mirrors the repayment of debt with interest: the longer the lag persists, the more costly and disruptive the resolution becomes.

Table~\ref{tab:debt-vs-lag} compares the two constructs across key dimensions. The table highlights that while TD arises from deliberate shortcuts, TL emerges passively through inaction. Despite this difference in origin, both produce debt-like liabilities with compounding costs, reduced quality, and increased risk.

\begin{table}[ht]
\footnotesize
\centering
\caption{Comparison of Technical Debt and Technical Lag}\label{tab:debt-vs-lag}

\begin{tabular}{p{2cm} p{2.5cm} p{2.9cm}}
\toprule
\textbf{Dimension} & \textbf{Technical Debt} & \textbf{Technical Lag} \\
\midrule
\textbf{Origin} & Intentional shortcuts or trade-offs in design or implementation & Passive accumulation from outdated dependencies, APIs, or platforms \\
\midrule
\textbf{Visibility} & Often identifiable through code smells, static analysis, or architectural violations & Often invisible without ecosystem-level metrics and tooling \\
\midrule
\textbf{Accumulation} & Results from conscious choices (e.g., deferring refactoring) & Results from inaction and inertia (e.g., not updating dependencies) \\
\midrule
\textbf{Cost Dynamics} & ``Interest'' accrues as defects, fragility, and maintenance effort & ``Compound Interest'' accrues as larger migrations, compatibility issues, dependency deprecation \\
\midrule
\textbf{Impact} & Reduced maintainability, higher defect rates, slower development velocity & Increased vulnerability exposure, fragile update processes, and ecosystem-wide risks \\
\midrule
\textbf{Management} & Refactoring, redesign, debt tracking, repayment planning & Automated dependency updates, ecosystem monitoring, lag-aware tooling \\
\midrule
\textbf{Conceptual Position} & Explicit, intentional liability & Latent, passive liability a hidden form of debt \\
\bottomrule
\end{tabular}
\end{table}

\section{Threats to Validity}\label{sec:threatsToValidity}


\subsubsection{Internal Validity}
Selection bias may impact internal validity, as our initial search was limited to studies that specifically mentioned TL and software ecosystems. This could exclude relevant studies that may be valuable but were not identified due to limited information on how TL was measured, or they lacked the correct keywords. Additionally, the snowballing process may have introduced bias by relying on citations from the initial set of studies, potentially overlooking less-cited but relevant work.

\subsubsection{External validity} Given that most studies heavily focus on ecosystems such as \texttt{npm} and \texttt{Debian}, our conclusions might not fully apply to less-studied ecosystems or niche technological environments. Therefore, caution is warranted when generalizing findings beyond the contexts explored in existing literature.

\subsubsection{Construct validity} Variations in how TL metrics were applied and modifications due to ecosystem challenges could impact our groupings. For example, some Time Lag metrics were dependent on the semantic versioning of the project and could have been classified as Version Lag, rather than Time Lag. Additionally, findings that equate newer with safer may overstate risk; recent high-profile supply-chain attacks show the need for risk-aware, signal-informed lag metrics.

\section{Conclusion}\label{sec:conclusion}

As summarized in this review, reframing TL as LTD gives organizations another tool to monitor software health. By positioning TL as a quantifiable contributor to TD, organizations can incorporate it into existing debt monitoring frameworks, creating a unified approach to managing both intentional and unintentional liabilities.
\balance
\bibliographystyle{ACM-Reference-Format}
\bibliography{paper}
\end{document}